\documentclass[twocolumn,showpacs,preprintnumbers,aps,prl,amsmath,amssymb]{revtex4}

\usepackage{graphicx}
\usepackage{dcolumn}
\usepackage{bm}
\usepackage{amssymb}
\usepackage[german, english]{babel}
\begin{document}
\draft

\title{Finite-Temperature Collective Dynamics of a Fermi Gas in the BEC-BCS Crossover}

\author{M. J. Wright,$^{1}$ S. Riedl,$^{1,2}$ A. Altmeyer,$^{1,2}$ C. Kohstall,$^{1}$
E. R. S$\mathrm{\acute{a}}$nchez Guajardo,$^{1}$\\ J. {Hecker
Denschlag},$^{1}$ and R. Grimm$^{1,2}$}

\address{$^{1}$Institut f\"ur Experimentalphysik und Forschungszentrum f\"ur Quantenphysik,
Universit\"at Innsbruck, 6020 Innsbruck, Austria\\
$^{2}$Institut f\"ur Quantenoptik und Quanteninformation,
\"Osterreichische Akademie der Wissenschaften, 6020 Innsbruck,
Austria}

\date{\today}


\begin{abstract}
We report on experimental studies on the collective behavior of a
strongly interacting Fermi gas with tunable interactions and
variable temperature. A scissors mode excitation in an elliptical
trap is used to characterize the dynamics of the quantum gas in
terms of hydrodynamic or near-collisionless behavior. We obtain a
crossover phase diagram for collisional properties, showing a
large region where a non-superfluid strongly interacting gas shows
hydrodynamic behavior. In a narrow interaction regime on the BCS
side of the crossover, we find a novel temperature-dependent
damping peak, suggesting a relation to the superfluid phase
transition.
\end{abstract}

\maketitle

Ultracold Fermi gases with tunable interactions have opened up
intriguing possibilities to study the crossover from bosonic to
fermionic behavior in strongly interacting many-body quantum
systems \cite{Inguscio2006ufg}. In the zero-temperature limit, a
Bose-Einstein condensate (BEC) of molecules is smoothly connected
with a superfluid of paired fermions in the
Bardeen-Cooper-Schrieffer (BCS) regime. In recent years, great
progress has been achieved in the theoretical description of the
ground state at zero temperature, and fundamental properties have
been experimentally tested with considerable accuracy
\cite{Giorgini2007tou}. Finite-temperature phenomena in the
BEC-BCS crossover, however, pose great challenges for their
theoretical description. Experimental observations of
finite-temperature behavior in the crossover have focussed on the
measurement of condensate fractions
\cite{Regal2004oor,Zwierlein2004cop}, on the spectroscopic
investigation of pairing phenomena \cite{Chin2004oop}, or on the
special case of unitarity-limited interactions
\cite{Kinast2005doa,Stewart2006peo,Clancy2007ooa}.

To understand the collective dynamics of an ultracold quantum gas,
it is crucial to study the conditions for hydrodynamic behavior.
Collective mode experiments have probed the dynamics of strongly
interacting Fermi gases for variable interaction strength near
zero temperature \cite{Bartenstein2004ceo, Kinast2004boh,
Altmeyer2007pmo, Altmeyer2007doa}. The results show the existence
of both a hydrodynamic regime of collective motion and a
near-collisionless regime with independent motion of the trapped
particles. The role of temperature, however, remained essentially
unexplored.

In this Letter, we explore the collective behavior of a
finite-temperature, strongly interacting Fermi gas of $^6$Li atoms
throughout the BEC-BCS crossover. In order to characterize the
transition from hydrodynamic to collisionless behavior, we analyze
scissors mode oscillations at different temperatures. With varying
temperature, the oscillations show a smooth transition between the
two collisional regimes along with a broad maximum in the damping
rate. We discover an unexpected second peak in the damping rate at
lower temperatures in a narrow region on the BCS side of the
crossover, where the gas remains hydrodynamic. This suggests the
lower-temperature damping peak to be connected to the transition
from a superfluid to a normal hydrodynamic gas.

\begin{figure}
\centerline{\includegraphics[width=7.5cm]{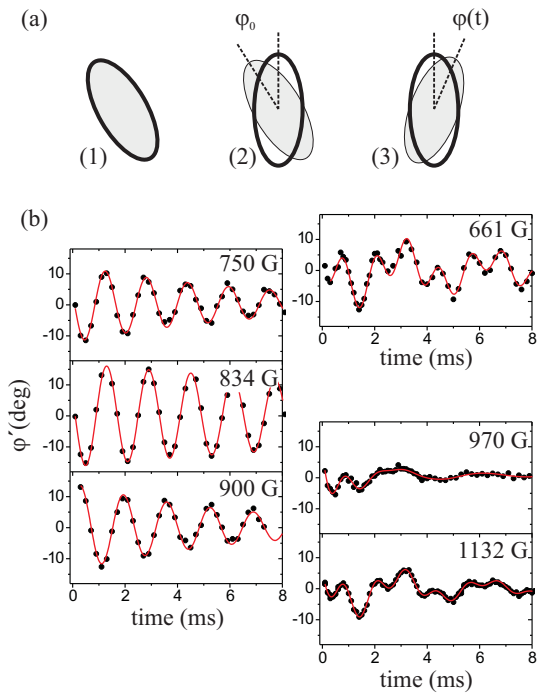}}
\caption{\label{lowTosc} (a) Schematic showing the excitation of
the scissors mode. (1) The gas (shaded region) is at rest, in
equilibrium with the trap (heavy solid line). (2) The trap is
suddenly rotated. (3) The gas oscillates around the new
equilibrium position. (b) Scissors mode oscillations observed at
the lowest obtainable temperature ($T \approx 0.1T_F$ at 834 G)
for various magnetic fields. On the left side, where $B$ $=$ 750
G, 834 G, and 900 G ($1/k_Fa = 1.4$, $0.0$, and $-0.6$), the gas
is hydrodynamic. On the right side, where $B = 661$ G, 970 G, and
1132 G ($1/k_Fa = 5.0$, $-1.0$, and $-1.44$), the gas is nearly
collisionless and exhibits the characteristic two-frequency
oscillation. Here $\omega_x$ = $2\pi \times 580$\,Hz, $\omega_y$ =
$2\pi \times 270$\,Hz, and $T_F = 0.69\,\mu$K.}
\end{figure}

The scissors mode in ultracold quantum gas experiments
\cite{Gueryodelin1999sma, Marago2000oot} is an angular oscillation
of the cloud about a principle axis of an elliptical trap, see
Fig.\ \ref{lowTosc}(a). In our experiments, we confine the atoms
in a harmonic, triaxial optical dipole trap. We choose the
geometry of the trap to produce an elliptically shaped gas in the
x-y plane with very weak confinement along the z axis. The
scissors mode experiments are done in the x-y plane. In terms of
trap frequencies, the standard configuration is $\omega_x >
\omega_y \gg \omega_z$, where these trap frequencies are defined
along the principle axes of the trap. If the gas is hydrodynamic,
the angle of the gas oscillates collectively with a single
frequency of $(\omega_x^2 + \omega_y^2)^{1/2}$. If the gas is
collisionless, the trapped atoms oscillate independently,
resulting in a two-frequency oscillation. The larger frequency is
given by $\omega_x + \omega_y$. When the collisional regime is
changed this frequency is adiabatically connected to the
hydrodynamic frequency. The smaller frequency is given by
$\omega_x - \omega_y$ and is absent in the hydrodynamic limit
\cite{Gueryodelin1999sma}.

The preparation of a strongly interacting Fermi gas of $^6$Li
proceeds in the same way as described in our previous work
\cite{Jochim2003bec,Altmeyer2007doa}. The result is a deeply
degenerate, balanced two-component spin mixture of typically $N =
4 \times 10^5$ atoms with tunable $s$-wave interactions near a
broad Feshbach resonance, which is centered at a magnetic field $B
= 834$\,G. Rapid spatial modulation of the trapping beam by two
acousto-optical deflectors is used to create a time-averaged
elliptical trapping potential for the scissors mode
\cite{Altmeyer2007doa}. The aspect ratio is set to
$\omega_x/\omega_y \approx 2.0$. We employ a trap with frequencies
$\omega_x= 2\pi \times 830$~Hz and $\omega_y = 2\pi \times 415$~Hz
($\omega_z$ = $2\pi \times 22$\,Hz), if not indicated otherwise.
This results in a Fermi temperature $T_F = (\hbar
\bar{\omega}/k_B)\,(3N)^{1/3} = 0.94 \mu$K, where
$\bar{\omega}=(\omega_x\omega_y\omega_z)^{1/3}$. The trap depth
corresponds to about 12\,$T_F$. To excite the scissors mode, we
suddenly rotate the angle of the trap by $\sim$ 5 degrees, see
Fig.\ 1(a).

The angle of the oscillating cloud is determined by processing
absorption images, taken after a short expansion time of
400\,$\mu$s. A two-dimensional Thomas-Fermi profile is fit to the
images, where the tilt of the principle axes of the cloud is a
free parameter, see Fig.\ \ref{lowTosc}(a). Note that the short
expansion somewhat decreases the ellipticity of the cloud, but
increases the amplitude of the scissors mode oscillation
\cite{Modugno2003smo}. In the hydrodynamic regime, we fit a damped
cosine function to the experimental data. In the collisionless
regime, we fit the oscillation to a sum of two damped cosine
functions each with their own free parameters. In the region
between these two limits, we find that a single damped cosine
function fits the data reasonably well, as the lower-frequency
component damps out very quickly \cite{Gueryodelin1999sma}.

First, we examine the collective behavior of the gas at our lowest
obtainable temperatures.  We compare scissors mode oscillations at
different settings of the magnetic field, i.e.\ different values
of $1/k_Fa$. Typical scissors mode oscillations are shown in Fig.\
\ref{lowTosc}(b). At $B = 661$ G, far on the BEC side of
resonance, the gas exhibits nearly collisionless behavior. Here
inelastic collisions result in heating the gas above the critical
temperature for BEC. In the regime where the gas is strongly
interacting, $B = 750$~G, 834~G, and 900~G, the gas oscillates
collectively. High precision measurements taken at $B = 834$ G
show the scissors mode oscillation yields a frequency that agrees
with theory within one percent. Far on the BCS side, at $B$ = 970
G and 1132 G, the gas exhibits behavior that is nearly
collisionless. The abrupt transition between the hydrodynamic and
collisionless regimes at low temperature occurs at essentially the
same magnetic field, $B$ $\approx$ 950 G, as in other collective
mode experiments
\cite{Bartenstein2004ceo,Kinast2004boh,Altmeyer2007doa}.

\begin{figure}
\centerline{\includegraphics[width=7.5cm]{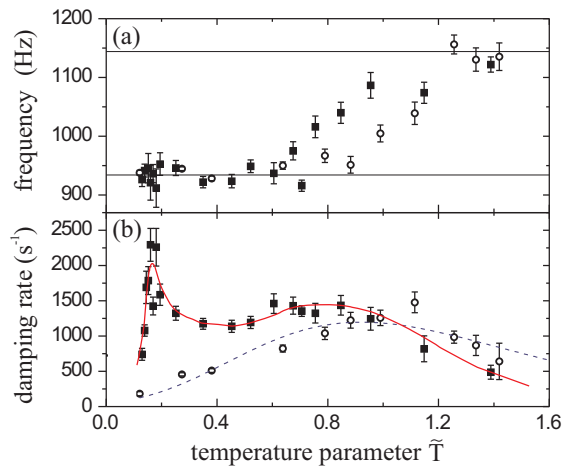}}
\caption{\label{freqvsT}Frequency and damping rate for the
scissors mode oscillation for $B$ = 895 G (1/$k_Fa$ = $-0.45$,
solid squares) and at $B$ = 834 G (1/$k_Fa$ = 0, open circles).
The frequency limits in the hydrodynamic and collisionless regimes
are shown by the horizontal lines in (a), including small
corrections for the anharmonicity of the trap \cite{Riedl07}. The
lines in (b) are introduced as guides to the eye. For $\tilde{T}$
greater than $1.14$, the scissors mode oscillations are fit by a
two-frequency cosine function (for details see text).}
\end{figure}

To explore finite-temperature collisional behavior, we extend the
scissors mode measurements. To set the temperature, we use a
controlled heating scheme. Here, we suddenly compress the trap and
allow for subsequent equilibration \cite{Axial_mode}. We control
the temperature of the gas by adjusting the amount of compression.

The determination of the temperature $T$ in an ultracold, strongly
interacting Fermi gas is in general difficult
\cite{Kinast2005hco}. We can measure an effective temperature (or
entropy) parameter $\tilde{T}$ at the center of the Feshbach
resonance, $B = 834$ G. We determine $\tilde{T}$ by fitting
integrated, one-dimensional, density profiles in the manner
described in \cite{Kinast2005hco, KinastPhD, Stajic2005dpo}. At $B
= 834$ G, for $T/T_F > 0.3$, the parameter $\tilde{T}$ is
proportional to the real temperature with $T/T_F \approx
\tilde{T}/1.5$. For lower temperatures, an empirical conversion
has been determined \cite{Kinast2005hco, KinastPhD, Stajic2005dpo,
temperature_eq}. The parameter $\tilde{T}$, measured in the
unitarity limit at 834\,G, can be used also as a temperature scale
for other interaction regimes under the condition that entropy is
conserved in adiabatic sweeps of the magnetic field
\cite{Chen2005toi}.

In Figure \ref{freqvsT}, we show the frequency and damping rate as
a function of $\tilde{T}$ for two cases, in the unitarity limit
($1/k_Fa = 0.00$) and at the BCS side of the crossover ($1/k_Fa =
-0.45$). The frequency behavior in Fig.~\ref{freqvsT}(a) is
qualitatively the same for both cases. At low temperatures, the
gas shows the hydrodynamic frequency and, at the highest
temperatures, we observe the behavior characteristic for the
collisionless gas. With varying temperature, the changing
frequency smoothly connects the hydrodynamic with the
collisionless regime. Quantitatively, the transition occurs at
somewhat higher $\tilde{T}$ in the unitarity limit. In the
transition region, the damping rate shows a maximum that
accompanies the change in frequency, see Fig.~\ref{freqvsT}(b). We
introduce the temperature parameter $\tilde{T}_H$ for this damping
maximum, marking the transition between hydrodynamic and
collisionless behavior.

The temperature dependence of the damping rate in Fig.\
\ref{freqvsT}(b) reveals a qualitatively different behavior between
the two interaction regimes. An additional peak shows up at lower
temperatures for the BCS side of the crossover, while this peak is
absent in the unitarity limit. Remarkably, this novel feature is not
associated with a change in the frequency.

\begin{figure}
\centerline{\includegraphics[width=6cm]{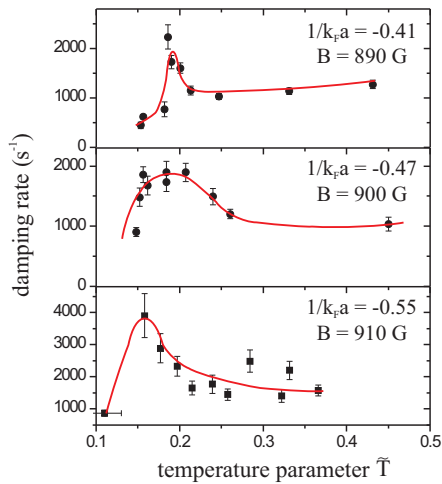}}
\caption{\label{smallpeak}Low-temperature damping peak observed in a
narrow magnetic-field region at the BCS side of the resonance
($1/k_Fa \approx -0.5$). The solid lines are introduced as guides to
the eye.}
\end{figure}

We could detect the low-temperature damping peak only in a very
narrow range at the BCS side of the crossover. This feature was
found between magnetic fields of 890\,G and 920\,G, corresponding to
interaction parameters $1/k_Fa$ between $-0.6$ and $-0.4$. In Fig.\
\ref{smallpeak}, we show the low-temperature damping peak as it
changes in this narrow region. Closer to resonance, the peak becomes
very narrow, shifts toward higher temperatures, and finally seems to
disappear. To mark the location of this peak, we introduce the
temperature parameter $\tilde{T}_S$.

\begin{figure}
\centerline{\includegraphics[width=7.5cm]{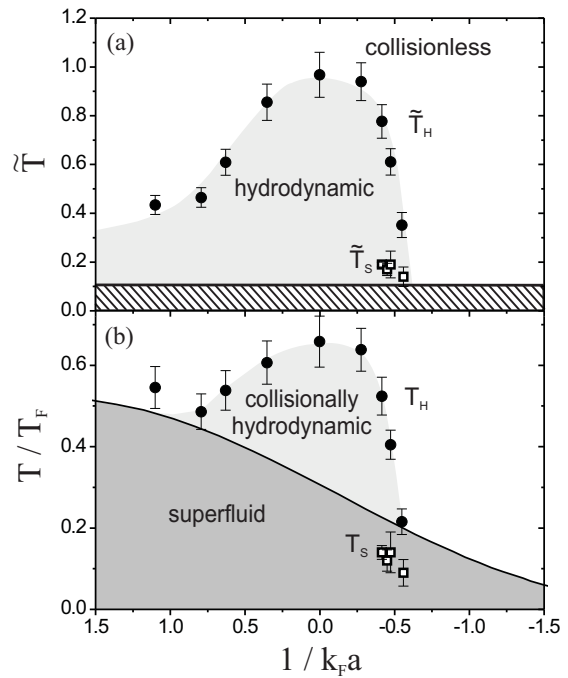}}
\caption{\label{phasediagram}Phase diagram for the hydrodynamic
behavior of the scissors mode in terms of (a) the temperature
parameter $\tilde{T}$ and (b) the real temperature $T$. The smooth
transition from hydrodynamic to collisionless is characterized by
the temperature parameter $\tilde{T}_H$ (temperature $T_H$). The
second damping peak near $1/k_Fa$ $\approx$ $-0.5$ is marked by
$\tilde{T}_S$ ($T_S$). In (a) the hatched region indicates the
region ($\tilde{T} < 0.1$) where our thermometry does not produce
reliable results. In (b) the solid line shows a theoretical curve
for the phase transition to superfluidity \cite{Perali2004bbc}.}
\end{figure}

We now discuss our observations in terms of a crossover phase
diagram for the scissors mode excitation \cite{compare}. In Fig.\
\ref{phasediagram}(a), we plot $\tilde{T}_H$ (closed circles) and
$\tilde{T}_S$ (open squares) as a function of the interaction
parameter. The data points for $\tilde{T}_H$ show a pronounced
maximum at the center of the resonance.
To facilitate an interpretation
of the experimental data, we convert $\tilde{T}_H$ and
$\tilde{T}_S$ into real temperatures $T_H$ and $T_S$, following
the theory of Ref.~\cite{Chen2005toi}. Fig.\ \ref{phasediagram}(b)
shows the resulting phase diagram, including a theoretical
prediction \cite{Perali2004bbc} of the temperature $T_C$ for the
phase transition to a superfluid state.

Near the center of the Feshbach resonance, hydrodynamic behavior
is observed far above the superfluid transition temperature. The
large difference between $T_H$ and $T_C$ confirms the existence of
a non-superfluid hydrodynamic region above $T_C$
\cite{Minguzzi2001smi,Kinast2005doa,Clancy2007ooa}. Our
measurements show that this normal-gas hydrodynamic regime is
restricted to the narrow, strongly interacting region near
resonance where $1/k_Fa$ stays well below unity. On the BEC side,
$T_H$ is close to the expected value for $T_C$. Here one can
assume that hydrodynamic behavior essentially results from the
formation of a molecular BEC. A surrounding non-condensed
molecular gas would exhibit near-collisionless properties, similar
to what has been measured in atomic BEC experiments
\cite{Marago2001tdo}. On the BCS side of resonance, $T_H$ falls
off very rapidly. In this region, collective modes may couple to
the weakly bound fermion pairs \cite{Bartenstein2004ceo,
Altmeyer2007doa}. We did not observe hydrodynamic behavior beyond
that point.

For the low-temperature damping peak found at the BCS side of the
crossover near $1/k_Fa\approx-0.5$, our phase diagram in
Fig.~\ref{phasediagram}(b) suggests a close relation to the
superfluid phase transition. The peak occurs at roughly
$0.6\,T_C$, and it follows the general behavior of the superfluid
transition to move toward higher temperature as it approaches the
resonance. This points to a scenario where a substantial
superfluid core in the center of the trap is surrounded by a
non-superfluid, but still hydrodynamic fraction in the outer
region of the trap. Whether damping results from the coupling of
these two components or whether other mechanisms are responsible
for this phenomenon remains an open question. We note that the
low-temperature damping peak is not specific to the scissors mode.
We have also found a corresponding, but less pronounced peak in
measurements of the radial breathing mode. Further investigations
and better theoretical understanding will be required to answer
the intriguing question whether the novel damping peak does indeed
mark the transition from the normal hydrodynamic to the superfluid
state.

In conclusion, we have investigated hydrodynamic behavior at
finite temperatures in the BEC-BCS crossover using scissors mode
excitations. Our measurements highlight the existence of a region
of non-superfluid hydrodynamics in the strongly interacting regime
where $|k_Fa| \gtrsim 1$. In the unitarity limit, predominant
hydrodynamic behavior is found up to $\sim$$0.6\,T_F$, which
substantially exceeds the superfluid transition temperature of
$\sim$$0.3\,T_F$. With increasing temperature, the transition from
hydrodynamic to collisionless behavior proceeds in general
smoothly and is accompanied by a local maximum of damping. In
addition, we have discovered a novel low-temperature damping peak
at the BCS side of the crossover, which suggests a relation to the
superfluid phase transition. With this observation, experiments on
collective oscillation modes of Fermi gases in the BEC-BCS
crossover continue to produce puzzling observations
\cite{Bartenstein2004ceo,Kinast2005doa,Altmeyer2007doa} with the
potential to stimulate deeper theoretical understanding of the
physics of strongly interacting Fermi gases.

We thank S.\ Stringari for stimulating discussions. We also thank
Q.\ Chen, K.\ Levin, and J. E.\ Thomas for discussions concerning
determination of the temperature. We acknowledge support by the
Austrian Science Fund (FWF) within SFB 15 (project part 21). M.J.W.\
is supported by a Marie Curie Incoming International Fellowship
within the 6th European Community Framework Program.


\begin{thebibliography}{27}
\bibitem[{Axi()}]{Axial_mode}
\bibinfo{note}{Sudden compression of the trap excites the axial mode which is
  long lived. Since the frequency is much smaller than the scissors mode
  frequency, it can be neglected. Nonetheless, we carried out a direct
  comparison without the axial mode present and found the same behavior.}
\end{thebibliography}

\end{document}